# Kilobyte-Scale, Selector-Free, Temperature-Hard AlScN Ferroelectric Diode Crossbar Arrays

Zirun Han[1,4†], Chao-Chuan Chen[1†], Dhiren K. Pradhan[1], David C. Moore[2], Ravali Gudavalli[3], Xindi Yang[3], Kwan-Ho Kim[1], Hyunmin Cho[1], Zachary Anderson[1], Spencer Ware[1], Harsh Yellai[1], W. Joshua Kennedy[2], Nicholas R. Glavin[2], Roy H. Olsson III[1], and Deep Jariwala[1,3]*

[1]*Department of Electrical and Systems Engineering, University of Pennsylvania, Philadelphia, Pennsylvania 19104, United States*

[2]*Materials and Manufacturing Directorate, Air Force Research Laboratory, Wright-Patterson AFB, Dayton, Ohio 45433, United States*

[3]*Department of Materials Science and Engineering, University of Pennsylvania, Philadelphia, Pennsylvania 19104, United States*

[4]*Department of Physics and Astronomy, University of Pennsylvania, Philadelphia, Pennsylvania 19104, United States*

[†] *These authors equally contributed to this work.*

*Author to whom correspondence should be addressed: dmj@seas.upenn.edu

**ABSTRACT**

We report the fabrication and characterization of kilobyte-scale, selector-free, ferroelectric (FE) diode crossbar memory arrays based on aluminum scandium nitride (AlScN). Utilizing a fully CMOS back-end-of-line (BEOL) compatible process, we fabricated 2-kilobyte (128×128) arrays with device diameters down to 5 µm, achieving memory densities up to 2500 bits/mm². Large-scale electrical characterization across 1000 randomly selected devices reveals a yield rate of 95.2%, a tight switching voltage distribution with a coefficient of variation (CV) of 0.003, and consistent on/off ratios of around 10 with a CV of 0.27. We demonstrate selector-free read and program operations of the array, enabled by the high nonlinearity, rectification, and uniform switching behavior of the FE diodes. Furthermore, we verified consistent ferroelectric switching during array operation at temperatures up to 600 °C. Our results highlight the potential of AlScN FE diode arrays for energy-efficient, high-density memory applications and lay the groundwork for future integration in compute-near-memory, high-temperature memory, and analog compute-in-memory systems.

**INTRODUCTION**

Modern data-driven computing applications, such as generative artificial intelligence (AI), are becoming increasingly memory-intensive due to the exponential scaling of model sizes. Traditional memory technologies, however, present significant bottlenecks due to the dual requirements of rapid data access for inference computations and nonvolatile storage of model weights. In the current paradigm, data and parameters are stored in slower, nonvolatile memory, such as flash memory, and subsequently loaded into faster, volatile memories like dynamic and static random-access memory (DRAM/SRAM) during computation. This data transfer is both slow and energy-inefficient, with each memory technology encountering distinct scaling challenges[1].

Potential solutions to address these challenges include integrating energy-efficient, nonvolatile memory directly above logic processors in a monolithic 3D stack, enabling compute-near-memory (CNM) architectures, or employing devices that integrate logic and memory functions entirely within compute-in-memory (CIM) architectures. Ferroelectric memory has emerged as a promising candidate for these applications due to its energy-efficient switching, inherent nonvolatility, and multistate programmability[2]. In particular, ferroelectric diodes (FE diodes) exhibit rectification behavior and significant nonlinearity, allowing them to effectively suppress sneak currents and enabling their integration into high-density crossbar arrays without additional selectors.

The appeal of FE diode memory is amplified by discovery of modern ferroelectric materials such as hafnium oxide[3] and aluminum scandium nitride (AlScN)[4]. Specifically, AlScN demonstrates uniform ferroelectric properties, excellent retention, and compatibility with CMOS back-end-of-line (BEOL) processes due to its low deposition temperature. While recent studies[2,5,6] have showcased excellent scaling and electrical properties of AlScN-based FE diodes using individual devices, the integration of these diodes within large-scale crossbar arrays remains unexplored.

In this work, we present a 2-kilobyte FE diode crossbar array memory utilizing ferroelectric AlScN, which marks a step from single-device demonstrations toward full memory technology integration. The arrays are fabricated using a scalable, fully CMOS BEOL-compatible process in a university cleanroom. We conducted extensive testing on 1000 randomly selected devices, revealing a high device yield rate of 95.2% and consistent electrical performance. Additionally, we demonstrate selector-free read and programming operations, highlighting the potential of this memory array for high-density, energy-efficient integration.

Additionally, the high Curie temperature of AlScN (> 1100 °C[7]) positions the AlScN-based FE diode as a strong candidate for high-temperature nonvolatile memory applications. The combination of selector-free operation and high-temperature compatibility is especially advantageous, as it significantly simplifies the complexity of the peripheral circuitry required to operate the memory at elevated temperatures. In comparison, other potential high-temperature memories, such as nanogap-based resistive memory[8], do not support selector-free operation. Previously, stable operation of single FE diodes at temperatures up to 600 °C has been reported[6]. Here, we demonstrate that the crossbar arrays we report, which are fabricated with temperature-hard materials, exhibit reliable ferroelectric switching and above-unity on/off ratios at 600 °C. These findings represent meaningful progress toward large-scale nonvolatile memory systems for extreme temperature applications.

**RESULTS AND DISCUSSION**
**<u>Structure of the AlScN FE Diode Crossbar Array</u>**

The memory presented in this work consists of 2-kilobyte, 128×128 crossbar arrays comprising 16,384 FE diodes. We demonstrate two scaled versions of this array. The first version, which we will refer to as the "10 µm array", features FE diodes with a diameter of 10 µm. The corresponding top and bottom electrode traces are 20 µm wide with a pitch of 40 µm, achieving a memory density of 625 bits/mm². Additionally, we fabricated a scaled-down version, referred to as the "5 µm array", with 5 µm diameter FE diodes, 10 µm wide electrode traces, and a pitch of 20 µm. This array achieves a higher memory density of 2500 bits/mm². While this is significantly lower than the state-of-the-art memory densities reported for other emerging memory technologies, such as ~15 gigabits/mm² for phase change memory (PCM)[9] and ~7 gigabits/mm² for resistive random access memory (RRAM)[10], the fabrication process described here supports further lateral scaling of FE diodes using advanced lithography techniques. For example, reducing the FE diode diameter to 50 nm, achievable with deep ultraviolet (DUV) or electron-beam lithography, would increase memory density to 0.1 gigabits/mm². Investigating the effects of lateral scaling on AlScN FE diodes will be the subject of a future study.

The FE diodes in the arrays are fabricated by first sputtering a 100 nm Pt bottom electrode (BE) with a 10 nm Ti adhesion layer onto a sapphire substrate. Then, a 45 nm ferroelectric $Al_{0.64}Sc_{0.36}N$ (AlScN) film is sputtered onto the Pt BE. To introduce asymmetry and improve the on/off ratio of the device, a 10 nm $HfO_x$ or $AlO_x$ interlayer (IL) is grown via atomic layer deposition (ALD) on top of the AlScN. A 270 nm $SiN_x$ passivation layer is then deposited via plasma-enhanced chemical vapor deposition (PECVD). The top electrode (TE), comprising a stack of 250 nm Cr and 100 nm Au, is deposited via electron beam evaporation and contacts the device through a via structure etched into the $SiN_x$ passivation layer. The entire fabrication process maintains a maximum temperature of 350 °C and ensures full BEOL compatibility for future integration onto a viable Si CMOS logic technology. Figure 1(a) provides a cross-sectional schematic along with images of the crossbar array, and the fabrication process is described in further detail in the Methods section. The process demonstrated here is scalable, as demonstrated by fabrication of the 5 µm arrays on a full 4-inch wafer as shown in Figure 1(e).

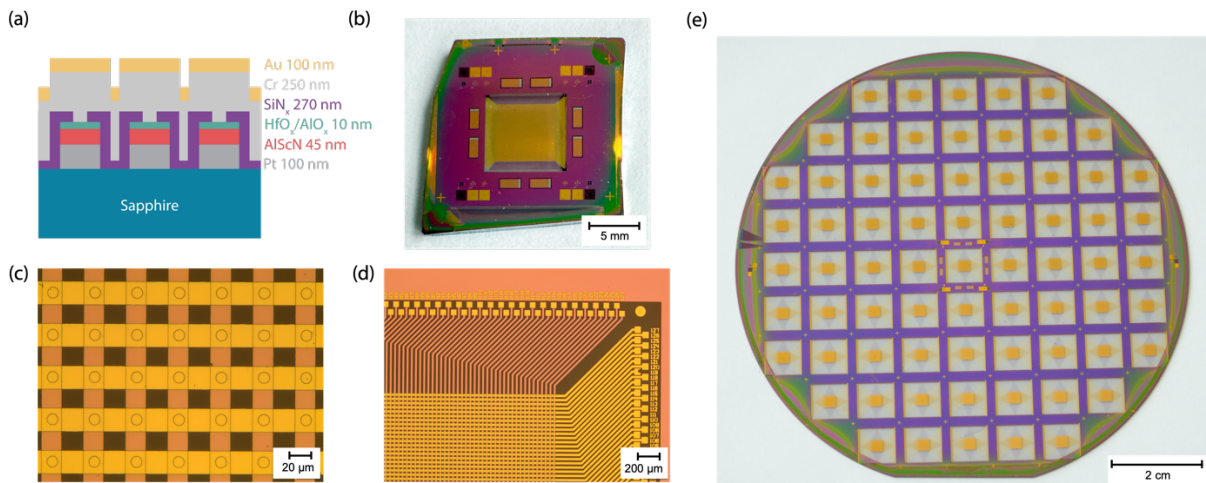

**Figure 1. Structure of the AlScN ferroelectric diode crossbar array. (a)** Cross-sectional diagram of the crossbar array structure. The illustration is not to scale. **(b)** Camera image of the 10 µm crossbar array used for large-scale electrical testing. The peripheral structures are alignment markers for photolithography and (non-crossbar) arrays of single FE diodes for diagnostic purposes. **(c)** Microscope image of the 10 µm crossbar array in (a) under 50x magnification. **(d)** Microscope image of the top right

section of the 10 μm crossbar array in (a) under 5x magnification. **(e)** Camera image of 5 μm crossbar arrays fabricated on a 4-inch wafer.

## Statistical Analysis of FE Diode Electrical Properties

To examine the consistency of the FE diodes in the array, we randomly selected 1000 devices with a Python script for electrical characterization. A 10 μm array with HfO$_x$ IL was used for this large-scale testing. The results of the measurements and associated statistical analysis are shown in Figure 2. In our measurement procedure, we apply the driving voltage and ground to the BE and TE, respectively, addressing the device under test while all other electrodes are left floating.

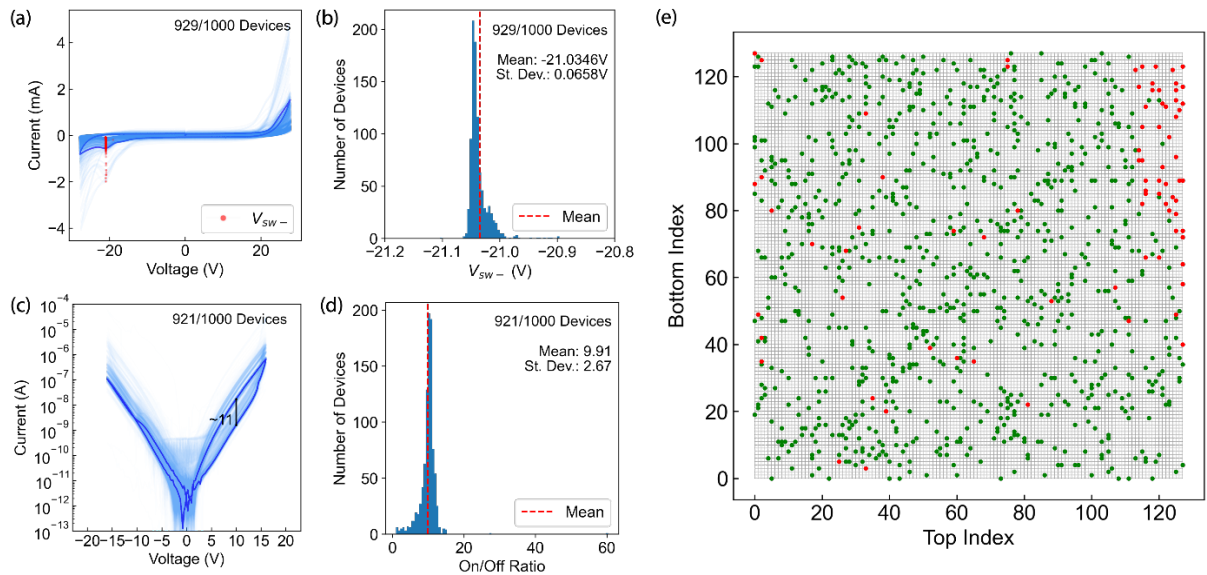

**Figure 2. Statistics of electrical characteristics of FE diodes in the crossbar array. (a)** AC I-V curves of the FE diodes in the 10 μm crossbar array with HfO$_x$ IL. The current response to a 12.5 kHz triangle wave is recorded. The negative switching voltage (V$_{sw-}$) of each device is extracted and labeled on the plot. A representative curve is highlighted in purple. The graph depicts the 929/1000 devices that survived the test at the maximum amplitude of 28 V. **(b)** Histogram plot of V$_{sw-}$ extracted from the data shown in (a). **(c)** DC I-V curves of the FE diodes. The current response to a quasi-static voltage sweep is recorded. A representative curve is highlighted in purple with its on/off ratio at 10 V labeled. The graph depicts the 921/1000 devices that survived the AC I-V test depicted in (a) and the DC I-V at the maximum amplitude of 16 V. **(d)** Histogram plot of on/off ratio at 10 V extracted from (c). **(e)** A map of the 1000 devices that were randomly tested. The devices that survived both the AC and DC I-V tests at maximum amplitude are shown in green, and devices that failed before the completion of the test sequence are shown in red.

Two tests are performed sequentially on each device. First, an AC current-voltage (I-V) measurement was performed, where 12.5 kHz triangle waves with amplitudes of 15 V, 20 V, 25 V, 26 V, 27 V, and 28 V were sequentially applied to the device and the current response is recorded. The main purpose of this measurement is to confirm ferroelectric switching and extract the switching voltage ($V_{sw}$) of the device, which is identified by a characteristic current peak. Although the switching peak is present for both polarities, the subsequent statistical analysis will focus on the negative switching voltage ($V_{sw-}$). This is because the negative-side leakage current is lower, resulting in a more clearly defined peak and enables more consistent extraction over a large number of devices. Figure 2(a) presents the AC I-V curves at the maximum voltage amplitude of 28 V. The locations of $V_{sw-}$ are found by applying a peak-finding function implementing a continuous wavelet transformation to the I-V data and labeled on the figure. Figure 2(b) summarizes the statistical distribution of $V_{sw-}$, highlighting a tight distribution with a mean of -21.0346 V and a standard deviation of 65.8 mV, resulting in a coefficient of variation (CV) of 0.003. This result compares favorably with other emerging memory technologies, like RRAM, where reported CV in switching voltage range from 0.03 to 0.70[11–13]. Our findings are also consistent with the observation of steep and uniform switching behavior in wurtzite ferroelectrics[14].

After the AC I-V measurement, a subsequent DC I-V measurement was used to extract the on/off ratio of the devices. For this test, a quasistatic voltage sweep with the segments [0 V, $V_{max}$], [$V_{max}$, -$V_{max}$], and [-$V_{max}$, 0V] is applied to the device under test. The value of $V_{max}$ was increased from 8 V, 12 V, 14 V, to 16 V in four sequential measurements. The results of the measurements at $V_{max}$ = 16 V are shown in Figure 2(c).

The on/off ratio of the device is defined to be the ratio of currents between the low resistance state (LRS) and the high resistance state (HRS). The statistical distribution of FE diode on/off ratios extracted at 10 V is shown in the histogram in Figure 2(d). The devices exhibit a mean on/off ratio of 9.91 with a standard deviation of 2.67. The on/off ratio of these devices is consistent with previously reported 45 nm AlScN devices[6], but lower than that of scaled (5 – 20 nm) FE diodes[15]. This is in part due to the reduced modulation of the electric field in the IL caused by ferroelectric polarization reversal when a thicker ferroelectric layer is used[15].

The distribution of the array's on/off ratio has a higher variation compared to the distribution of its switching voltage ($V_{sw-}$). The tight distribution of switching voltage indicates that the array is largely ready for application as capacitor-based ferroelectric random-access memory (FeRAM), while the larger variation in on/off ratio indicates that material and design improvements are required for successful application as a resistive memory. We note that while the switching voltage is primarily dictated by the properties of the ferroelectric layer, the on/off ratio is largely determined by the quality and dielectric properties of the IL[15]. As such, improving the uniformity and quality of the IL is an immediate next step for the optimization of these devices. Furthermore, the problem of on/off ratio variation can be mitigated by scaling down the thickness of the ferroelectric layer to 10 – 20 nm. This leads to an increase of the on/off ratio, which means variation in the on/off ratio is less likely to affect state distinguishability.

Overall, 929 out of 1000 devices withstood the AC I-V test at 28 V, and 921 devices subsequently survived the DC I-V test at 16 V, achieving a survival rate of 92.1%. A map of the device failures is shown in Figure 2(e). Notably, our measurements stress the devices beyond $V_{sw}$ by a significant margin. The maximum AC voltage of 28 V used in our measurement is 33.2% higher than the measured AC $V_{sw}$ of 21.03 V, and the maximum DC voltage of 16 V is 14.3% higher than the DC $V_{sw}$ of approximately 14 V. In a practical application of the crossbar array, the devices will likely be operated much closer to their switching voltages. As such, we can define device yield based on successful ferroelectric switching (at 25 V AC conditions), which results in a yield rate of 95.2% (952 out of 1000 devices). The spatial distribution of the failed devices is mostly uniform other than a block of devices in the top right of the array that had a very high failure rate, which is likely due to a microfabrication and handling induced defect rather than an issue with intrinsic material uniformity. Continued material and process optimizations are expected to further enhance these metrics especially when ported into a commercial foundry-based process.

**Selector-free Operation of the FE Diode Array**

The ability to operate a memristive crossbar array without dedicated selector transistors is an advantageous property as it allows for simpler integration and more energy-efficient operation[16]. Here, we demonstrate both selector-free read and program operations of the FE diode crossbar arrays using the floating scheme, which leaves the unselected TEs and BEs floating. Notably, this is more energy-efficient and simpler to implement compared to the V/2 and V/3 schemes commonly used for other memristive crossbar arrays[16]. Selector-free current readout in the crossbar arrays presented in this work is enabled by the high nonlinearity and rectification behavior of the FE diodes. In the floating scheme, the shortest possible sneak path consists of three devices connected in series, each experiencing a fraction of the applied voltage. As such, the intrinsic nonlinearity of the FE diodes substantially suppresses currents through these sneak paths. Additionally, we leverage the diode's rectification behavior to further limit sneak currents, as all sneak paths necessarily include at least one FE diode under reverse bias. A graphical illustration of these concepts is provided in Figure S1 of the Supplemental Information.

Selector-free reading of the devices is confirmed by the DC I-V curves shown in Figure 2(c), which were measured using the selector-free floating scheme. As shown in Figure 2(c), the HRS and LRS of each device is clearly distinguishable with an appreciable on/off ratio in this setup. The soundness of this approach is further verified in Figure S2 in the supplemental information, where we compare the selector-free DC I-V characteristics of FE diodes in the crossbar array to that of a discrete device, where the array devices show only marginally higher current density due to sneak paths.

Selector-free programming of devices in the crossbar array requires a uniform switching voltage distribution of the FE diodes, which was demonstrated in Figure 3(b). Ferroelectric switching is a capacitive process that is driven by the potential difference applied across the device. Again, three FE diodes divide the programming voltage in the shortest sneak path. The exact voltage division depends on the frequency of the switching signal – in the low-frequency regime, the resistive characteristics of the devices would largely determine the voltage drop across each device, while the capacitive characteristics would dominate in the high-frequency regime. In any case, we expect a tight distribution of switching voltage (as shown for our FE diodes in Figure 2 a-b) to be important for selector-free programming as it prevents the unwanted switching of the neighboring devices when subject to a fraction of the applied switching voltage.

(a)

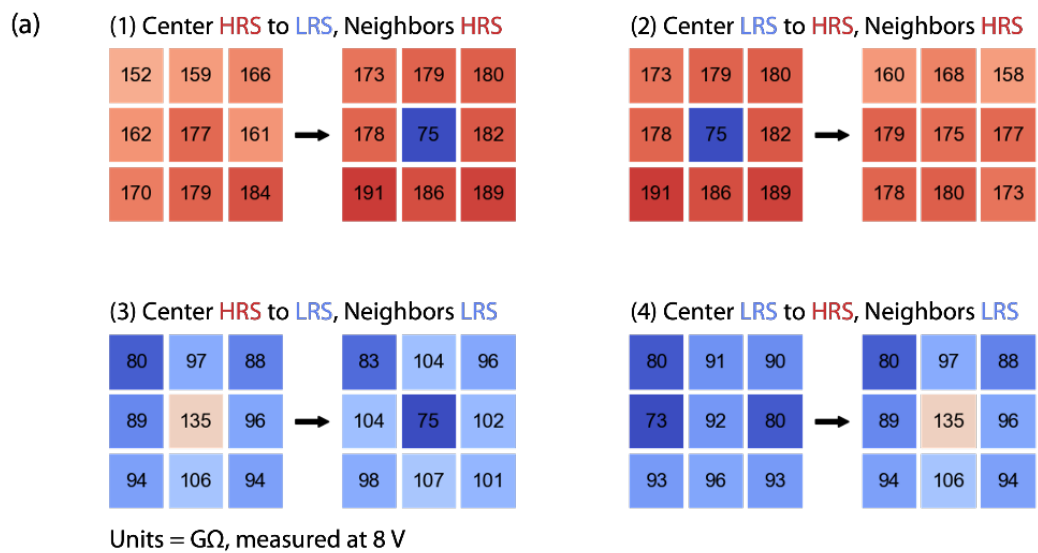

(1) Center HRS to LRS, Neighbors HRS

| 152 | 159 | 166 |
| 162 | 177 | 161 |
| 170 | 179 | 184 |

→

| 173 | 179 | 180 |
| 178 | 75 | 182 |
| 191 | 186 | 189 |

(2) Center LRS to HRS, Neighbors HRS

| 173 | 179 | 180 |
| 178 | 75 | 182 |
| 191 | 186 | 189 |

→

| 160 | 168 | 158 |
| 179 | 175 | 177 |
| 178 | 180 | 173 |

(3) Center HRS to LRS, Neighbors LRS

| 80 | 97 | 88 |
| 89 | 135 | 96 |
| 94 | 106 | 94 |

→

| 83 | 104 | 96 |
| 104 | 75 | 102 |
| 98 | 107 | 101 |

(4) Center LRS to HRS, Neighbors LRS

| 80 | 91 | 90 |
| 73 | 92 | 80 |
| 93 | 96 | 93 |

→

| 80 | 97 | 88 |
| 89 | 135 | 96 |
| 94 | 106 | 94 |

Units = GΩ, measured at 8 V

(b)

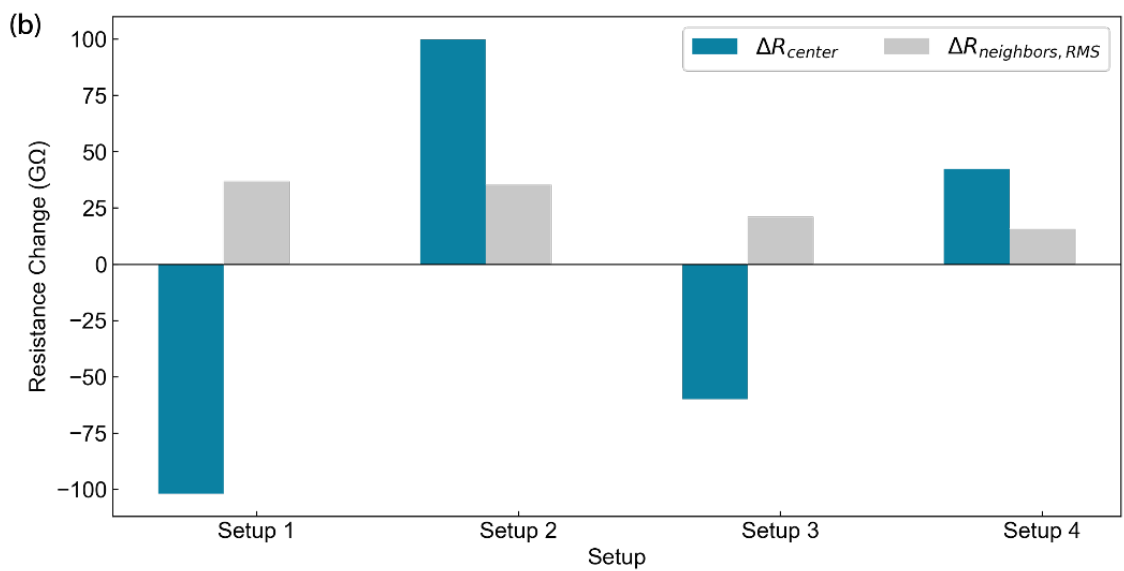

**Figure 3. Selector-free programming of the crossbar array. (a)** Demonstration of selector-free switching behavior in four scenarios: (1) switching the center device from HRS to LRS while neighboring devices remain in HRS; (2) switching the center device from LRS to HRS, neighbors remain in HRS; (3) switching the center device from HRS to LRS, neighbors remain in LRS; (4) switching the center device from LRS to HRS, neighbors remain in LRS. The values shown represent the measured resistance (in GΩ) at an 8 V read voltage. Colors indicate the resistance state on a blue-red spectrum, with blue representing LRS and red representing HRS. **(b)** Change in resistance (ΔR) of the center device and the root-mean-square (RMS) average change in resistance in neighboring devices for each setup. An RMS average is used here to prevent resistance changes in opposite directions cancelling each other.

To demonstrate the selector-free programming of the crossbar array, we performed ferroelectric switching on a central FE diode and monitored the resistive states of its immediate neighbors. Because failed devices result in a short between a TE and BE, they reduce the number of devices in series in the sneak path and adversely impact the results of this measurement. For this reason, a pristine 5 µm array is used. Using a DC switching signal, we performed both the LRS to HRS and HRS to LRS switching of the central device while the neighbors are in LRS and HRS, totaling four setup combinations. We measured the resistance of the devices at 8 V before and after the central device switching, and the results for each of the four setups are shown in Figure 3, where the change in resistance (ΔR) for the central device and the root-mean-square (RMS) average of ΔR for neighboring devices are compared for each of the four setups. A detailed procedure on how the devices were switched and measured is available in the supplemental information.

For setups 1 and 2, which has HRS neighbors, $\Delta R_{RMS}$ of the central device is 101 GΩ while $\Delta R_{RMS}$ of the neighboring devices is 36 GΩ. For setups 3 and 4, which has LRS neighbors, $\Delta R_{RMS}$ of the central device is 51 GΩ while $\Delta R_{RMS}$ of the neighboring devices is 18.5 GΩ. The change in resistance of the central device in the case of LRS neighbors is notably lower than the case of HRS neighbors. However, we note that this is likely not caused by the improper switching of the central FE diode but is rather due to the lower sneak path resistance in the LRS setup reducing the measured resistance of the central device in HRS during the subsequent read operation. Across the four setups, a minimum margin of 28 GΩ is maintained between the HRS and LRS. Although further improvements are needed to achieve fully stable operation, our results demonstrate the potential for AlScN FE diode crossbar arrays to support selector-free programming.

**High-Temperature Operation of the FE Diode Crossbar Array**

To demonstrate the suitability of AlScN FE diode crossbar arrays as high-temperature nonvolatile memory, we performed electrical characterization of a 10 µm array featuring a 10 nm $AlO_x$ interlayer at 600 °C. $AlO_x$ was chosen as the interlayer material due to its high crystallization temperature of around 800 °C[17] , making it suitable for high-temperature applications. The characterization included selector-free DC-IV and AC-IV sweeps similar to the room-temperature tests, alongside an endurance test using positive-up-negative-down (PUND) pulse sequences applied until device breakdown. Detailed high-temperature measurement protocols are provided in the Supplemental Information. Of the 23 sampled devices, 21 successfully completed the entire test procedure without premature failure, with results illustrated in Figure 4.

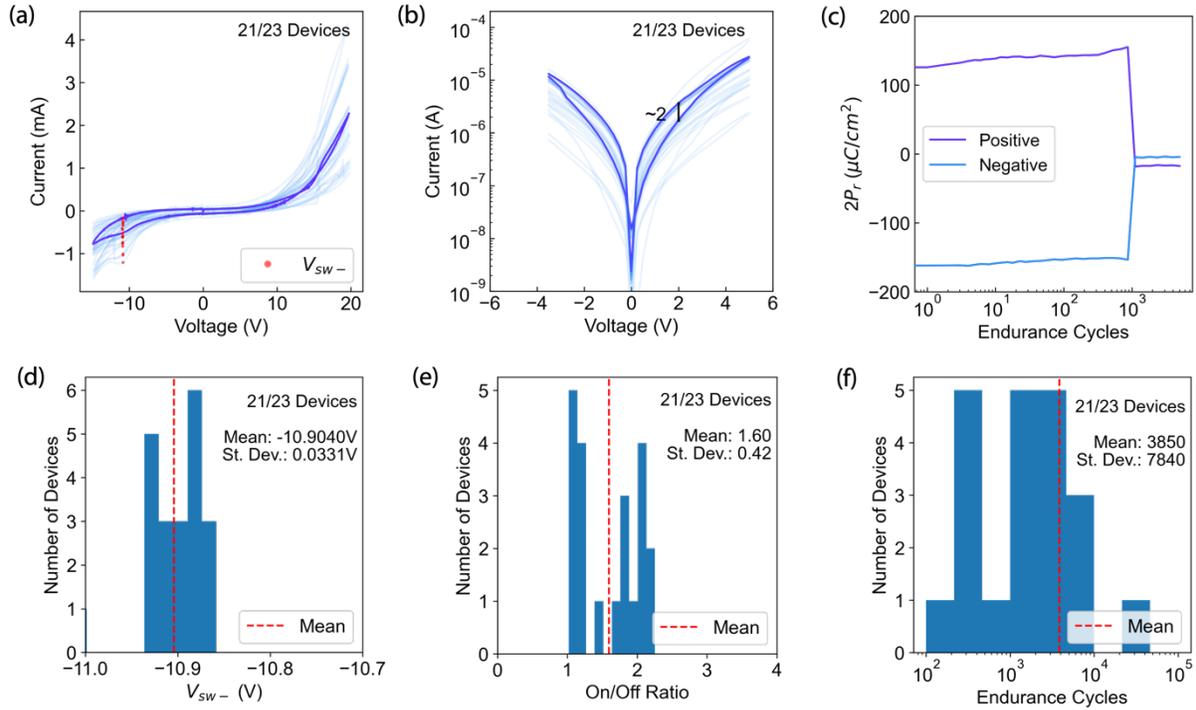

**Figure 4. 600 °C Operation of the FE Diode Array. (a)** AC-IV curves of FE diodes in the 10 μm crossbar array with AlO$_x$ IL. Current response to a 12.5 kHz triangle wave is recorded and a representative curve is highlighted. **(b)** DC-IV curves of the FE diodes. Current response to the quasi-static voltage sweep is recorded and a representative curve is highlighted. **(c)** The remnant polarization of a representative device measured periodically during PUND endurance testing. **(d)** Histogram plot of V$_{sw-}$ extracted from the AC-IV data in (a). **(e)** Histogram plot of On/Off ratios measured at 2V extracted from the DC-IV data in (b). **(f)** Histogram plot of device endurance using PUND stress cycles shown in (c).

AC-IV measurements (Figure 4(a)) demonstrate that the devices maintain a tight switching voltage distribution at elevated temperatures, as summarized in Figure 4(d). DC-IV measurements at 600 °C (Figure 4(b)) confirmed above-unity on/off ratios but also revealed increased variability, as depicted in Figure 4(e). While uniform switching characteristics affirm the suitability of these FE diodes for capacitive memory applications, the observed variability in their I-V characteristics currently limits their effectiveness for current-readout memory applications. Improving the consistency of the on/off ratios and I-V characteristics will be the subject of future research.

Since device endurance typically deteriorates at elevated temperatures[6], we conducted endurance tests using the PUND pulses at the conclusion of the characterization sequence, with results shown in Figure 4(c). An example of the PUND sequence used is shown in Figure S3 in the supplemental information. Endurance of the devices ranged from 129 to 37,785 cycles, detailed by the histogram in Figure 4(f). We note that endurance measurements were not performed at room temperature, as these tests inevitably lead to device breakdown that results in the shorting of the corresponding BE and TE, adversely impacting large-scale characterization.

**CONCLUSION**

In this work, we introduce 2 KB, BEOL-compatible FE diode crossbar memory arrays based on ferroelectric AlScN. We performed large-scale characterization of the device switching voltage and on/off ratio, reporting a device yield of 95.2% and uniform characteristics. Furthermore, we demonstrate selector-free read and write operations on the array and reliable ferroelectric switching at 600 °C. The memory array presented here holds significant potential for a broad range of applications, including BEOL memory in silicon processors, high-temperature memory, and analog memory.

To advance the research presented in this work, scaling down the device thickness for room temperature applications would decrease operating voltage and improve the on/off and rectification ratios, enhancing selector-free operation and suitability for resistive memory applications. Continued improvements in the quality of AlScN and the IL materials will be essential to further increase device yield and enhance I-V characteristic consistency. Furthermore, investigating lateral scaling of FE diodes into the sub-micron regime is critical for achieving bit densities competitive with conventional and other emerging memory technologies. Finally, exploring novel compute-in-memory applications that leverage the multistate polarization of AlScN, such as analog matrix-vector multiplication, offers a promising direction for future research.

## METHODS

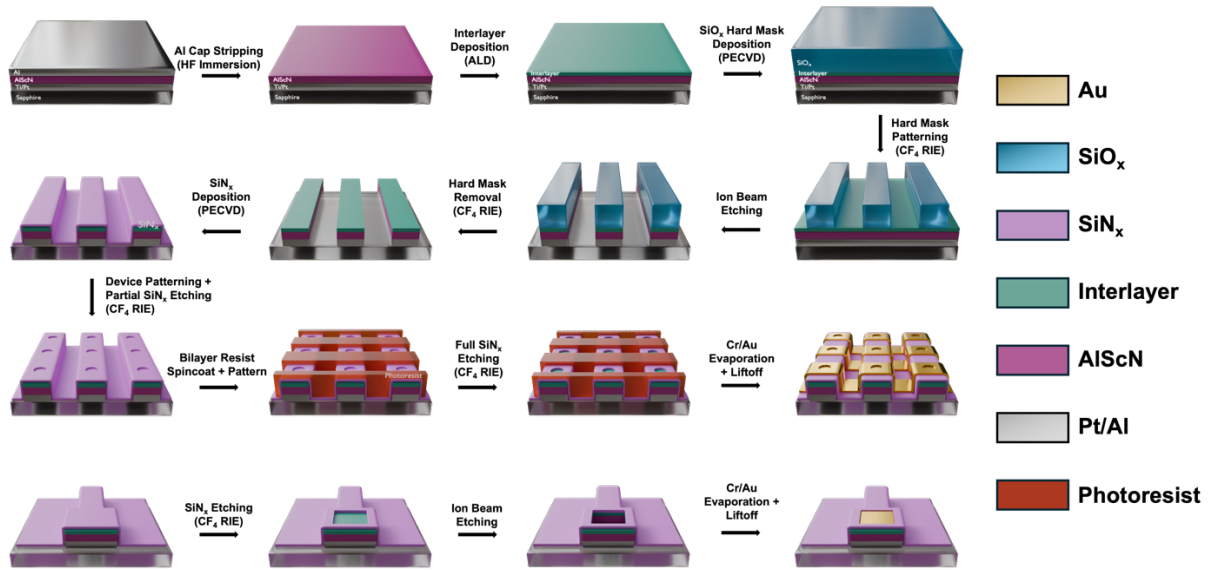

**Figure 5. Fabrication process flow for AlScN ferroelectric diode crossbar arrays.** A detailed procedure for the fabrication of the crossbar arrays is given in the device fabrication section.

### Array Layout Design

The layouts of the crossbar arrays are designed using a custom Python script developed using the open-source PHIDL library[18].

### AlScN Deposition

45 nm thick AlScN thin films with 36% Sc concentration ($Al_{0.64}Sc_{0.36}N$) were grown on Ti (10 nm)/Pt (100 nm) coated, c-plane sapphire wafers (both 4" and 6") using a physical vapor deposition (PVD) system. First, a 10 nm-thick Ti adhesion layer was deposited on the c-axis-oriented sapphire wafer by DC sputtering at 350 °C. Without breaking vacuum, a 100 nm-thick Pt layer was deposited at 350 °C. Then, ~ 45 nm of AlScN was deposited from separate Al and Sc targets, each 100 mm in diameter. The co-sputtering was carried out in an Evatec CLUSTERLINE® 200 II pulsed DC PVD system at substrate temperature of 350 °C. Following the growth of the AlScN film, a 50 nm-thick Al capping layer was deposited on top of the AlScN film. The Al capping layer was deposited in situ without breaking the vacuum to prevent surface oxidation of the AlScN film.

### Device Fabrication.

*1. Material Stack Preparation*: First, the protective Al layer on top of AlScN is removed by immersion in a 2% HF solution for 80 seconds immediately before array fabrication. Then, a 10 nm $HfO_x$ or $AlO_x$ dielectric layer is deposited at 250 °C by atomic layer deposition (ALD) using a Cambridge Nanotech S200 system.

*2. Bottom Electrode Patterning*: A 500 nm $SiO_x$ hard mask is grown on top of the stack through plasma-enhanced chemical vapor deposition (PECVD) using an Oxford Instruments PlasmaLab 100 system. The bottom electrode design is patterned with UV photolithography. The hard mask is then etched using a reactive ion etching (RIE) process on an Oxford Instruments 80Plus system. The unmasked parts of the sample are then etched down to the sapphire substrate through ion beam etching (IBE) using an Intlvac NanoQuest 1 system. The hard mask is subsequently removed using an RIE process. Then, a passivation layer consisting of 300 nm of $SiN_x$ is grown on top of the sample using PECVD.

*3. Device Patterning*: The array of devices is patterned using photolithography, and the passivation layer on top of the circular device areas is etched using RIE until 30 nm of $SiN_x$ remained. The purpose of this under etching is to use the $SiN_x$ layer to protect the devices from the tetramethylammonium hydroxide (TMAH)-based developer during the patterning of the top electrodes.

*4. Top Electrode Patterning*: The negative image of the top electrodes is patterned using photolithography. Then, 30 nm of the exposed $SiN_x$ layer is etched using RIE so that the nitride layer on top of the device is fully etched. Then, 250 nm Cr and 100 nm Au are sequentially deposited on the sample via electron beam evaporation without breaking vacuum using a Kurt J. Lesker PVD75 system. The sample is sonicated in a heated N-Methyl-2-Pyrrolidone (NMP) solution to complete the lift-off process.

*5. Bottom Pad Opening*: The bottom electrode pads are patterned using photolithography. The $SiN_x$ layer on top of the bottom electrode pads is etched using an RIE process, and the $HfO_x$ and AlScN layers are etched via IBE. A 50 nm Cr/ 100 nm Au layer is deposited on top of the exposed Pt layer through electron beam evaporation and subsequent lift-off. A graphical illustration of the fabrication process is shown in Figure 5.

**Electrical Characterization**

The electrical characterization is performed at room temperature using a Keithley 4200A-SCS semiconductor parameter analyzer connected to a precision probe station. High temperature measurements were performed in a custom MicroXact high-temperature probe station using tungsten probes, and electrical characterization was carried out with a Keithley 4200A-SCS parameter analyzer.

**ASSOCIATED CONTENT**

Supplemental information with Figure S1 illustrating sneak paths in the array during selector-free, floating-scheme operation, Figure S2 showing a comparison between selector-free DC I-V characteristics of array devices and a discrete device, and Figure S3 showing an example PUND sequence. A detailed description of the measurement procedure for selector-free programming of the array and high-temperature characterization are also provided.

**RESOURCE AVAILABILITY**

**Lead Contact**

Requests for further information and resources should be directed to and will be fulfilled by the lead contact, Deep Jariwala (dmj@seas.upenn.edu).

**Code Availability**

The Python code used to generate the crossbar array designs and the GDS files for the 10 µm and 5 µm 2 KB arrays, as well as the GDS files of the arrays fabricated, are provided in the following GitHub repository:

https://github.com/danny1078/crossbar_array_generator

**Material Availability**

This study did not generate new unique materials.

**DECLARATION OF INTERESTS**
The authors declare no competing interests.

**AUTHOR INFORMATION**

**Author Contributions**

Z.H. and C.-C.C. contributed equally to this work.

D.J., R.H.O., D.K.P., and Z.H. conceived the idea of AlScN-based ferroelectric diode crossbar arrays. Z.H. designed the array layout, wrote the code for array generation, and performed the device fabrication. D.K.P., K.-H.K., and H.Y. deposited the AlScN. C.-C.C., R.G., X.Y., Z.H., Z.A. and S.W. performed the electrical measurements. Z.H. and C.-C.C. performed the data analysis. H.C. performed microscopy during process optimization. D.C.M. performed the high temperature measurements under supervision of W.J.K and N.R.G. Z.H. wrote the manuscript with inputs from D.K.P. and C.-C.C. All authors read and revised the manuscript.

## ACKNOWLEDGEMENTS

Z.H. acknowledges funding support from the Vagelos Integrated Program in Energy Research (VIPER) at Penn. This material is based on research sponsored by Air Force Research Laboratory under agreement number FA8650-20-2-5506, as conducted through the flexible hybrid electronics manufacturing innovation institute, NextFlex. The U.S. Government is authorized to reproduce and distribute reprints for governmental purposes notwithstanding any copyright notation thereon. The views and conclusions contained herein are those of the authors and should not be interpreted as necessarily representing the official policies or endorsements, either expressed or implied, of Air Force Research Laboratory or the U.S. Government. A portion of the sample fabrication, assembly and characterization were carried out at the Singh Center for Nanotechnology at the University of Pennsylvania, which is supported by the National Science Foundation (NSF) National Nanotechnology Coordinated Infrastructure Program grant NNCI-1542153.

Supplemental Information

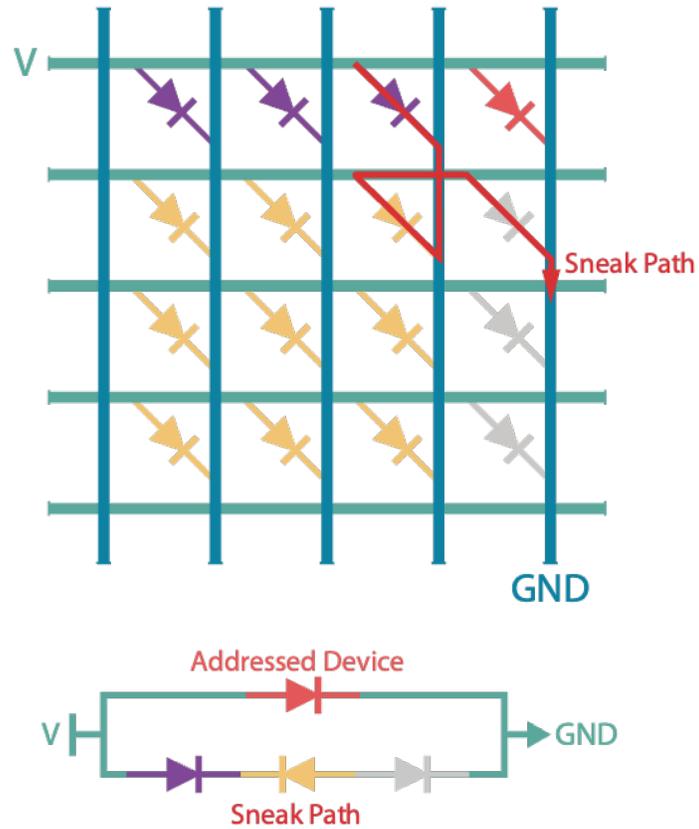

**Figure S1. Illustration of sneak paths in the crossbar array in the floating scheme.** In this diagram, the FE diode shown in light red is addressed by applying a program/read voltage and ground to its corresponding bottom electrode (BE) and top electrode (TE) respectively. All other lines are left floating. An example of the shortest sneak path possible in this configuration is labeled and shown in dark red. We note that any sneak path through the array must necessarily put a FE diode in the path in reverse bias. As such, the rectification behavior of the devices plays an important role in suppressing sneak currents and enabling selector-free operation.

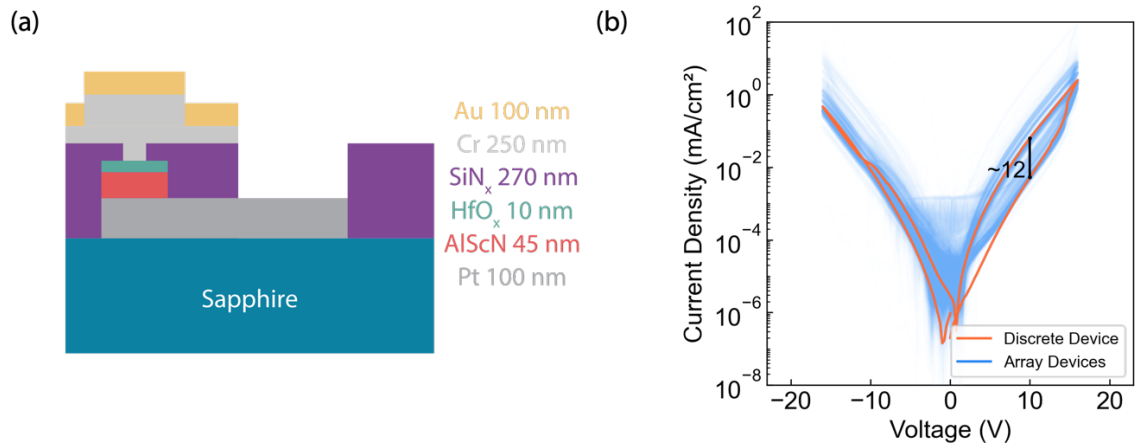

**Figure S2. Comparison of DC I-V characteristics between FE diodes in the crossbar array and discrete FE diodes. (a)** A cross-sectional diagram of a standalone FE diode. **(b)** The DC current density-voltage (J-V) characteristics of a 20 μm diameter discrete FE diode overlayed on the DC J-V curves of devices in the 10 μm crossbar array as presented in Figure 2(c) in the main manuscript. The on/off ratio of the standalone device is marked on the plot. The discrete diode is fabricated in close proximity to the crossbar array on the same chip. The J-V characteristics demonstrate the minimal difference in electrical characteristics between a discrete device and devices measured in the 2 KB crossbar array in a selector-free manner.

**Measurement Procedure for Testing Selector-Free Programming**

To verify that the devices in our array are selector-free, a 3×3 subset of devices from the array were selected from a pristine 5 μm array for measurement. To polarize the devices into the desired resistance states, a quasistatic DC voltage sweep from 0 V to 16 V and back to 0 V was used to set devices into the low resistance state (LRS), while a sweep from 0 V to 16 V, then to –16 V, and back to 0 V was used to set devices into the high resistance state (HRS). After each polarization step, the resistance of all 9 devices was measured at 8 V.

For setups 1 and 2 shown in Figure 3(a) in the main manuscript, all devices were first polarized into the HRS. The central device was then switched into the LRS (setup 1) and then returned to the HRS (setup 2). For setups 3 and 4, all devices were initially polarized into the LRS. The central device was then switched into the HRS (setup 4) and subsequently returned to the LRS setup (3).

**Measurement Procedure for 600 °C Operation**

600 °C measurements were performed in a custom MicroXact high-temperature probe station using tungsten probes, and electrical characterization was carried out with a Keithley 4200A-SCS parameter analyzer. Each device underwent a battery of tests in the following order: wake-up positive-up negative-down (PUND) pulses, DC–IV, AC–IV, and write endurance. The wake-up PUND consisted of three consecutive PUND pulse sequences with P and U pulse amplitudes of 19.5 V, N and D pulse amplitudes of -14.5 V, pulse widths of 5 μs, rise times of 500 ns, and pulse delays of 5 μs. A segment of PUND pulses is shown in Figure S3. DC–IV measurements were performed as a double sweep from -3.5 V to 5 V, while AC–IV measurements were conducted at 12.5 kHz sweeping from –15 V to 20 V. Endurance testing comprised repeated PUND sequences with the same configuration as the wake-up PUND. To quantify cycles endured, a series of unrecorded waveforms were applied before each read PUND, with the number of unrecorded intervening PUND pulses increasing logarithmically between reads.

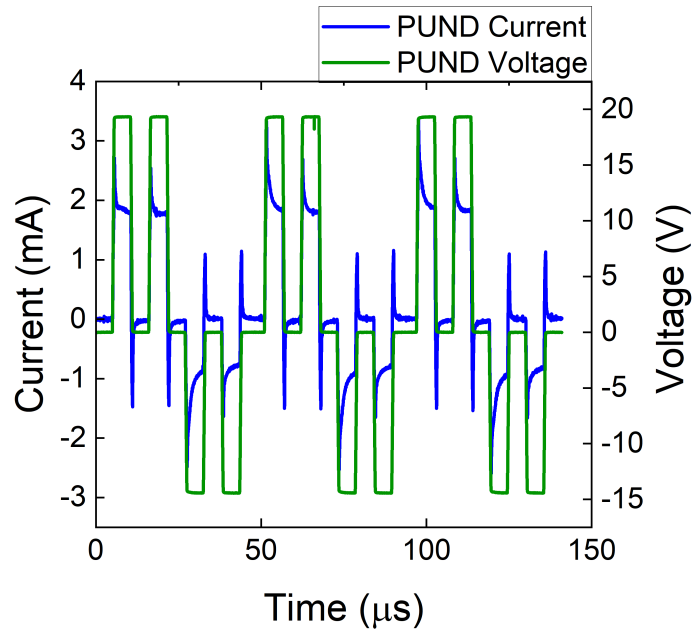

**Figure S3. Endurance Test PUND Sequences at 600 °C.** In PUND sequences, two consecutive pulses of the same polarity are applied to the device, where the first pulse (P/N) polarizes the ferroelectric and the second pulse (U/D) is used to account for leakage current in the device. The change in polarization is found by subtracting the current responses to the first and second voltage pulse and then integrating to find the accumulated polarization charge. Three PUND cycles, applied at 600 °C with the parameters described in the section above, are shown in this figure.